# Intrinsic Mobility Limiting Mechanisms in Lanthanum-doped Strontium Titanate

Amit Verma[1*], Adam P. Kajdos[2], Tyler A. Cain[2], Susanne Stemmer[2] and Debdeep Jena[1]

[1]Department of Electrical Engineering, University of Notre Dame, Notre Dame, Indiana 46556, U.S.A.
[2]Materials Department, University of California, Santa Barbara, California 93106, U.S.A.
*E-mail: averma@nd.edu

Abstract: The temperature dependent Hall mobility data from La-doped $SrTiO_3$ thin films has been analyzed and modeled considering various electron scattering mechanisms. We find that a ~6 meV transverse optical phonon (TO) deformation potential scattering mechanism is necessary to explain the dependence of transport on temperature between 10-200 K. Also, we find that the low temperature electron mobility in intrinsic (nominally undoped) $SrTiO_3$ is limited by acoustic phonon scattering. Adding the above two scattering mechanisms to longitudinal optical phonon (LO) and ionized impurity scattering mechanisms, excellent *quantitative* agreement between mobility measurement and model is achieved in the whole temperature range (2-300K) and carrier concentrations ranging over a few orders of magnitude ($8x10^{17}$ $cm^{-3}$ - $2x10^{20}$ $cm^{-3}$).

Understanding the carrier scattering mechanisms in any material paves the way for improvement in its transport properties, unraveling of new physics, and eventual applications. A good case in point is the semiconductor GaAs: improvements in the mobility of this material led to the discovery of the fractional quantum Hall-effect, and in ultrafast high electron mobility transistors (HEMTs) that enables high-speed microwave communications [1,2]. Strontium titanate, $SrTiO_3$, is a perovskite transition metal oxide. It has been studied extensively for the many interesting phenomena it exhibits, such as quantum paraelectricity, structural phase transition, superconductivity etc. [3-5]. The discovery of a two dimensional electron gas at the heterointerface of $SrTiO_3$ and other oxides has opened up a whole new area of heterostructure oxide electronics [6-8]. A major obstacle in developing devices using $SrTiO_3$ is the low electron mobility in this material, and a lack of complete understanding of the major active electron scattering mechanisms.

Electron transport studies in SrTiO$_3$ have been carried out for several decades, but without consensus [9-15]. There is general agreement that the dominant electron scattering mechanism at high temperatures is LO phonon scattering, and that ionized impurity scattering dominates at low temperatures, but no scattering mechanism has satisfactorily explained the mobility data at intermediate temperatures (~10-200K) [12,13,15]. A possible role of the ferroelectric lattice mode in explaining electron mobilities in SrTiO$_3$ at intermediate temperatures was suggested, but no confirmative and quantitative proof of the importance of this mode at intermediate temperatures has been established till date [10-13]. Kholkin et. al. [16] claimed to explain SrTiO$_3$ mobilities quantitatively at intermediate temperatures using a ferroelectric soft-phonon scattering mechanism [17]. Their results are, however, inconclusive, being based on a single highly defective sample with extremely low mobility (~10 cm$^2$/V-sec). Therefore, the question of the dominant electron scattering mechanism in SrTiO$_3$ at intermediate temperatures still remains unanswered. In most of the earlier transport studies, n-type doping was achieved either by the reduction of SrTiO$_3$ samples, or by introducing donor impurities (Nb, La) during single crystal growth [9-14]. The maximum low temperature mobilities achieved by these methods was ~22,000 cm$^2$/V-sec [11,15]. In recent years, the growth of SrTiO$_3$ by low energetic deposition techniques like molecular beam epitaxy (MBE) has enabled high quality thin films with carefully controlled doping and record high electron mobilities (more than 32,000 cm$^2$/V-sec) [18]. The recent availability of such high quality thin films enabled the study of the intrinsic mobility limits in SrTiO$_3$ in this work.

In this letter, we report on the dominant electron scattering mechanisms in La-doped SrTiO$_3$ in different temperature and carrier concentration ranges. Hall-effect mobility data from MBE grown SrTiO$_3$ thin films is used for exploring the transport limits [19]. At the low and high temperature ranges, we find that ionized impurity and polar LO phonons, respectively, are the dominant electron scattering mechanisms, in agreement with previous studies [11,12]. However, in the intermediate temperature (~10-200K) range, where no clear understanding of mobility limiting carrier scattering mechanism exists, we find that both temperature and carrier concentration dependence of measured mobility can be explained using a ~6 meV phonon deformation potential scattering mechanism. This phonon is quite possibly the TO soft-phonon mode associated with the antiferrodistortive phase transition of SrTiO$_3$ from cubic to the tetragonal phase [20]. We note that the soft mode associated with the antiferrodistortive transition is distinct from that associated with the ferroelectric transition studied earlier. We also find that the intrinsic low temperature mobility in SrTiO$_3$ is limited by acoustic phonon deformation potential scattering.

The experimentally measured mobility data used in this study is from samples grown by hybrid MBE. This low energetic technique uses high purity sources and has been demonstrated to produce stoichiometric SrTiO$_3$ films with record low temperature mobilities and highly controlled and precise La doping levels [18]. For the current study, La-doped SrTiO$_3$ thin films were grown on bulk (001) SrTiO$_3$ substrates. The electron concentration was varied over more than two orders of magnitude from $n$~8x10$^{17}$ cm$^{-3}$ to 2x10$^{20}$ cm$^{-3}$ by changing the La dopant concentration. Sheet resistance and Hall-effect measurements were performed in a Van der Pauw geometry over 2-300K temperature range. The Hall resistance $R_H = -1/ne$ where $e$ is the electron charge was found to be almost constant for each sample over the entire temperature range of measurement. This is expected because of the high dielectric constant of SrTiO$_3$ ($\varepsilon \sim 370-18000$ for 296-1.4 K [21]). From the hydrogenic model of donor impurities, the donor activation energy is inversely proportional to square of the dielectric constant of host material. A high dielectric constant of SrTiO$_3$ makes the La donor a very shallow impurity, which remains fully ionized even at 2K. The sample with $n$~8x10$^{17}$ cm$^{-3}$ doping has a mobility of ~53,000 cm$^2$/V-sec at 2K. This mobility value is highest ever reported for either bulk or unstrained thin film SrTiO$_3$, proving the high quality of these thin films and making them ideal for investigating intrinsic carrier scattering mechanisms in this material. Further details on the growth, processing and measurement of the samples can be found elsewhere [19].

Figure 1a presents the measured Hall mobility data as a function of temperature for all samples. This Hall data has been published previously [19] but is shown here again for completeness and for comparison with the modeling results. The calculated mobility fits accounting for scattering by two mechanisms, LO phonons and ionized impurities, for all samples are shown for comparison. The contribution due to the two individual scattering mechanisms is also shown separately for the lowest doping (8x10$^{17}$ cm$^{-3}$) sample in this figure. The calculated curves follow the treatment by Frederikse et. al. [12]. For the LO phonon scattering, a *polaron* model as derived by Low and Pines [22] is used according to which mobility variation is given as [12],

$$\mu_{LO} = \frac{\hbar}{2\alpha \hbar \omega_l} \frac{e}{m_P} \left(\frac{m_e^*}{m_P}\right)^2 f(\alpha)\left(e^{\hbar \omega_l / k_B T} - 1\right), \qquad (1)$$

where $e$ is electron charge, $\hbar = h/2\pi$ is the reduced Planck's constant, $\hbar \omega_l$ is the energy of the LO phonon mode involved in scattering, $\alpha$ is the electron-phonon coupling constant, $m_e^* = 1.8 m_e$ is the electron effective mass [23] where $m_e$ is the free electron mass, $m_P = m_e^*(1 + \alpha/6)$ is the polaron mass, $f(\alpha)$ is a slowly varying function ranging from 1.0 to

1.4, $k_B$ is Boltzmann constant, and $T$ is the absolute temperature. In SrTiO$_3$ two LO phonon modes with energies, $\hbar\omega_1 = 99$ meV and $\hbar\omega_2 = 58$ meV, are involved in electron scattering. The corresponding electron-phonon coupling constants are $\alpha_1 = 2.6$ and $\alpha_2 = 0.7$ [12]. The mobility due to each LO phonon-mode is calculated using Eqn.1 and the reciprocals of the mobilities are then added [12] to obtain the net mobility shown in Fig.1 (a). In the literature, depending on the experimental method used, there is a large spread in reported values of electron effective mass in SrTiO$_3$. Many methods report a value between 1-2 $m_e$ [23-26]). For this work, we use an electron effective mass of $1.8 m_e$ for all scattering mechanisms (except ionized impurity scattering as discussed later) as this value gives a *quantitative* agreement of the calculated mobility with the experimentally measured values. We also point out that if we neglect the many-particle polaronic aspect of electron-LO phonon interaction and resort to the single-particle Frohlich perturbation method as is done for polar III-V semiconductors such as GaAs and GaN, the calculated mobility values are very similar to those measured experimentally for SrTiO$_3$.

For modeling the ionized impurity scattering, we have used results obtained using the partial wave expansion technique [12, 27], which is valid for $ka \ll 1$, where $k$ is the electron wave vector and $a$ is the screening radius. Because the carrier concentration in the SrTiO$_3$ samples is independent of temperature, $k = k_F$, where $k_F$ is the Fermi wave vector. The Thomas-Fermi expression for the screening radius is given by [12],

$$a_F = \left( \frac{\hbar^2 \varepsilon}{4 m_{II}^* e^2} \left( \frac{\pi}{3 n_{3D}} \right)^{1/3} \right)^{1/2} \qquad (2)$$

where, $\varepsilon$ is the dielectric constant, $n_{3D}$ is electron density, and $m_{II}^*$ is the effective mass for ionized impurity scattering [28]. According to the partial wave expansion method [12], the ionized-impurity limited mobility is given by

$$\mu_{II} = \frac{e(\gamma^2 a^2 + 20\gamma a + 32)}{64\pi N_i v_F m_{II}^* a^2 (\gamma^2 a^2 + 8\gamma a)}, \qquad (3)$$

where, $a = a_F$, $N_i = n_{3D}$ is identical to the impurity concentration because of complete ionization, $v_F = (2 E_F / m_{II}^*)^{1/2}$ is the electron velocity at the Fermi energy $E_F$, and

$\gamma = \left(2m_{II}^* e^2 / \hbar^2 \varepsilon\right)$. Whether the long-wavelength condition $ka \ll 1$ under which Eqn.3 is valid is satisfied or not depends on our choice of electron effective mass $m_{II}^*$ and the dielectric constant $\varepsilon$ because the screening radius $a$ depends on both these material parameters. We know that the low-frequency dielectric constant of SrTiO$_3$ increases rapidly at low temperatures [21]. If we choose such high dielectric constant values, then $ka \gg 1$ is satisfied. Under such a condition, the theoretical treatment of ionized impurity scattering by Conwell-Weisskopf and Brook-Herring widely used for traditional doped semiconductors should be valid [12]. But, when we calculate the mobilities using these methods, we obtain mobility numbers that over-estimate the experimentally measured values by more than two orders of magnitude. Such overestimation of low-temperature mobility when using formulas derived under the Born approximation has been reported earlier [11]. On the other hand, if one uses the high-frequency dielectric constant value of SrTiO$_3$ ($\varepsilon = 4\varepsilon_0$, $\varepsilon_0$ being free space permittivity) the opposite condition $ka \ll 1$ is satisfied. Then we must use the partial wave expansion technique because Born-approximation fails in this regime. In addition to using the high-frequency dielectric constant value, if we also use a higher effective mass value of $m_{II}^* = 6.5 m_e$, the calculated and measured low-temperature mobilities agree very well as shown in Fig.1b. We leave the question of whether the use of these material parameters are physical or not for future discussion. If we estimate the polaron radius [12], $r_P = \left(\hbar^2 / 2m_{II}^* \hbar \omega_l\right)^{1/2}$ in SrTiO$_3$, and the screening radius $a_F$, we find that $r_P > a_F$. This relationship suggests that when the electron "sees" the ionized impurity scattering potential, it is buried deep inside a highly distorted polaronic potential well, and the effective mass should be much larger compared to the undistorted band-edge effective mass of SrTiO$_3$. The correct effective mass for the electron in such strong perturbation, and the effective dielectric constant value in such a distorted region needs further theoretical analysis. Thus, the use of the particular effective mass and dielectric constant as fitting parameters for the excellent fit to experiments over various samples is not completely unreasonable, but merits a more careful theoretical treatment of the problem in the future.

It is clear from Fig.1a that if we include only the effects of LO phonons and ionized-impurities on SrTiO$_3$ electron mobility, the net calculated mobility is highly overestimated (nearly two orders of magnitude for the lowest doped sample at ~60K as shown by the grey arrow in Fig.1a) in a wide intermediate temperature window from ~10 – 200 K. This disagreement hints at the existence of additional scattering mechanisms at play in SrTiO$_3$. The acoustic phonon scattering limited electron mobility is given by [29],

$$\mu_{ac} = \frac{2e\hbar\rho v_s^2}{3\pi n_{3d} m_e^* a_C^2} \ln(1+e^{E_F/k_BT}), \qquad (4)$$

where, $\rho = 5120$ kg/m$^3$ is the mass density of SrTiO$_3$, $v_s = 7900$ m/s is the sound velocity [31], $a_C = 4$ eV is the conduction band deformation potential [32], and $E_F$ is the Fermi energy. The calculated mobility including the effect of acoustic phonons is shown in Fig. 2a. Clearly, the slope of mobility decrease with increasing temperature in the lowest temperature range from 2-5 K is explained by the inclusion of acoustic phonon scattering. This feature can be more easily observed in Fig 2b where the mobility values have been normalized with respect to the mobility at 2K. This correct explanation of mobility slope suggests the important role of acoustic phonons in limiting the low temperature mobility in pure SrTiO$_3$ crystals. Acoustic phonons set the low temperature (2K) mobility limit in intrinsic undeformed SrTiO$_3$ at ~2x10$^5$ cm$^2$/V-sec.

However, even after including the effect of scattering by LO phonons, ionized impurities, and acoustic phonons, there is still a large mismatch of the calculated mobility in the 10-200K temperature range as can be seen from Fig. 2a. For example, for the lowest doped sample, the calculated mobility is ~20X larger than the measured value at T~60K. The overestimation is temperature-dependent. If we assume that an additional phonon mode of energy $\hbar\omega_{ph}$ is the dominant electron scattering mechanism in this intermediate temperature regime, the electron mobility should vary as $\mu \propto [\exp(\hbar\omega_{ph}/k_BT)-1]$. A fit of this mobility expression to the measured mobility of the lightest doped sample with carrier concentration ~8x10$^{17}$ cm$^{-3}$ in the 20-80K temperature range suggests that a phonon with $\hbar\omega_{ph}$ ~ 6 meV is at play in SrTiO$_3$. While we come to this conclusion from the analysis of the transport data, such a phonon mode has been identified by other experimental means. A 5.95 meV soft TO phonon mode has been identified in SrTiO$_3$ and is responsible for the 105 K cubic to tetragonal structural phase transition in this material [20,33]. It is quite possible that this soft-TO phonon mode also limits electron mobilities in SrTiO$_3$ in the intermediate temperature window.

In contrast to polar scattering by LO phonons, the atomic motion in a TO phonon mode does not produce macroscopic electric fields. Therefore, the only way it can scatter the electrons is by acting as local microscopic distortions, shifting the conduction band locally [30]. The soft-TO phonon mode in SrTiO$_3$ might be acting via such a non-polar deformation potential scattering. To test this hypothesis, we calculate the electron mobility limited by TO phonon deformation potential scattering, given by [29]

$$\mu_{TO} = \frac{\sqrt{2}\pi e \hbar^3 \rho v_s^2}{(m_e^*)^{5/2} \omega_{TO} D_{lop}^2} \frac{(e^{\hbar\omega_{TO}/k_B T}-1)}{\sqrt{E+\hbar\omega_{TO}}}, \qquad (5)$$

where, $\hbar\omega_{TO} = 5.95$ meV is the TO phonon energy, $D_{lop}$ is the optical phonon deformation potential, and $E(=E_F)$ is the energy of electron. Since, $D_{lop}$ is not known for SrTiO$_3$, we estimate a reasonable value of ~19 eV for this parameter, as calculated from fitting Eqn.5 to the mobility data of ~8x10$^{17}$ cm$^{-3}$ doped sample in 20-80K temperature range. Also, as shown in Fig.3b (Inset), the soft-TO phonon mode has a temperature dependence $\omega_{TO} = C(T-T_0)^a$, where $C$ is a constant, $T_0 = 105$ K is the structural phase transition temperature of SrTiO$_3$, and $a = 0.31$ [20]. For mobility modeling purposes we assume a temperature independent constant TO phonon energy of 5.95 meV. This assumption does not affect the calculated mobilities significantly because in the limit $T \to T_0$, the TO phonon mode softens ($\omega_{TO} \to 0$) and $\mu_{TO}$ becomes independent of $\omega_{TO}$ as can be seen from eqn.5. All other material parameters are same as those used for acoustic phonon scattering. Figures 3a and 3b show the modeled mobility and conductivity respectively, with the inclusion of TO phonon deformation potential scattering. Excellent *quantitative* agreement between the calculation and the experimental values over the whole temperature range, for all samples with different carrier concentrations, suggests that the ~6 meV TO phonon deformation potential scattering is indeed responsible for limiting mobility at intermediate temperatures in SrTiO$_3$.

In the past, the question of weak coupling of TO phonon modes with conduction band electron states has been raised [12]. In traditional semiconductors, such behavior is observed because of the s-orbital like symmetry of the conduction band electron states. This band symmetry leads to zero or quite small TO phonon scattering matrix elements [29,30]. Valence band electron states in traditional semiconductors on the other hand have p-orbital like symmetry. Consequently TO phonons can interact strongly with holes (for example in Si, Ge and GaAs [29,30]). The conduction band electron states of SrTiO$_3$ are quite similar to the valence band of traditional semiconductors. The conduction band electron states have the symmetry of Ti t$_{2g}$ d-orbitals. This analogy with valence band of semiconductors suggests the important role TO phonons can play in SrTiO$_3$ electron scattering.

To summarize, we have shown that combining traditionally used LO phonon and ionized impurity scattering mechanism, with acoustic phonon and a ~6 meV TO phonon scattering, SrTiO$_3$ electron mobility variation with temperature and carrier concentration can be explained quantitatively. We have identified the important role played by the ~6 meV soft-TO phonon mode at intermediate temperatures, which resolves a large discrepancy between pre-existing theories and in light of new experimental data. The work also clarifies the role played by other scattering mechanisms in action in $SrTiO_3$ and their relative importance in different temperature and carrier concentration ranges. We hope this work will motivate improved theoretical models for ionized impurity scattering in this material in the presence of very strong perturbation, and simultaneously seed more studies to further our understanding of transport in $SrTiO_3$, ultimately leading to improvement of the electron mobility in this technologically important material.

The authors thank Santosh Raghavan for useful discussions. This work was supported by the Extreme Electron Concentration Devices (EXEDE) MURI program of the Office of Naval Research (ONR) through grant No. N00014-12-1-0976.

**References**:

[1] H. L. Stormer, Rev. Mod. Phys. **71**, 875 (1999).
[2] T. Mimura, S. Hiyamizu, T. Fujii, and K. Nanbu, Jpn. J. Appl. Phys. **19**, L225 (1980).
[3] K. A. Muller and H. Burkard, Phys. Rev. B **19**, 3593 (1979).
[4] Farrel W. Lytle, J. Appl. Phys. 35, 2212 (1964).
[5] J. F. Schooley, W. R. Hosler, and Marvin L. Cohen, Phys. Rev. Lett. **12**, 474 (1964).
[6] A. Ohmoto and H. Y. Hwang, Nature **427**, 423 (2004).
[7] P. Moetakef, T. A. Cain, D. G. Ouellette, J. Y. Zhang, D. O. Klenov, A. Janotti, C. G. Van de Walle, S. Rajan, S. J. Allen, and S. Stemmer, Appl. Phys. Lett. **99**, 232116 (2011).
[8] H. Y. Hwang, Y. Iwasa, M. Kawasaki, B. Keimer, N. Nagaosa, and Y. Tokura, Nature Materials **11**, 103 (2012).
[9] H. P. R. Frederikse, W. R. Thurber, and W. R. Hosler, Phys. Rev. **134**, A442 (1964).
[10] S. H. Wemple, A. Jayaraman, and M. DiDomenico, Jr., Phys. Rev. Lett. **17**, 142 (1966).
[11] O. N. Tufte and P. W. Chapman, Phys. Rev. **155**, 796 (1967).
[12] H. P. R. Frederikse and W. R. Hosler, Phys. Rev. **161**, 822 (1967).
[13] S. H. Wemple, M. DiDomenico, Jr., and A. Jayaraman, Phys. Rev. **180**, 547 (1969).
[14] C. Lee, J. Destry, and J. L. Brebner, Phys. Rev. B **11**, 2299 (1975).
[15] A. Spinelli, M. A. Torija, C. Liu, C. Jan, and C. Leighton, Phys. Rev. B **81**, 155110 (2010).
[16] A. L. Kholkin, E. V. Kuchis, and V. A. Trepakov, Ferroelectrics **83**, 135 (1988).
[17] Y. N. Epifanov, A. P. Levanyuk, and G. M. Levanyuk, Ferroelectrics **35**, 199 (1981).
[18] J. Son, P. Moutakef, B. Jalan, O. Bierwagen, N. J. Wright, R. Engel-Herbert, and S. Stemmer, Nature Materials **9**, 482 (2010).
[19] T. A. Cain, A. P. Kajdos, and S. Stemmer, Appl. Phys. Lett. **102**, 182101 (2013).
[20] P. A. Fleury, J. F. Scott, and J. M. Worlock, Phys. Rev. **21**, 16 (1968).
[21] H. E. Weaver, J. Phys. Chem. Solids **11**, 274 (1959).
[22] Francis E. Low and David Pines, Phys. Rev. **161**, 414 (1955).
[23] M. Ahrens, R. Merkle, B. Rahmati, and J. Maier, Physica B 393, 239 (2007).
[24] J. L. M. van Mechelen, D. van der Marel, C. Grimaldi, A. B. Kuzmenko, N. P. Armitage, N. Reyren, H. Hagemann, and I. I. Mazin, Phys. Rev. Lett. **100**, 226403 (2008).
[25] D. van der Marel, J. L. M. van Mechelen, and I. I. Mazin, Phys. Rev. B **84**, 205111 (2011).
[26] S. James Allen, B. Jalan, S. Lee, D. G. Ouellette, G. Khalsa, J. Jaroszynski, S. Stemmer, and A. H. MacDonald, Phys. Rev. B **88**, 045114 (2013).
[27] Yu V. Gulyaev, Fiz. Tver. Tela **1**, 422 (1959) [English transl.: Soviet Phys. – Solid State **1**, 381 (1959)].
[28] Expression for Thomas-Fermi screening length actually involves the density of states effective mass but to reduce the parameters involved, we are using a single effective mass in ionized impurity limited mobility calculation.
[29] C. Hamaguchi, *Basic Semiconductor Physics* (Second Edition, Springer, 2010).
[30] P. Y. Yu and M. Cardona, *Fundamentals of Semiconductors* (Fourth Edition, Springer, 2010).
[31] R. O. Bell and G. Rupprecht, Phys. Rev. **129**, 90 (1963).
[32] A. Janotti, B. Jalan, S. Stemmer, and C. G. Van de Walle, Appl. Phys. Lett. **100**, 262104 (2012).
[33] R. A. Evarestov, E. Blokhin, D. Gryaznov, E. A. Kotomin, and J. Maier, Phys. Rev. B **83**,

134108 (2011).

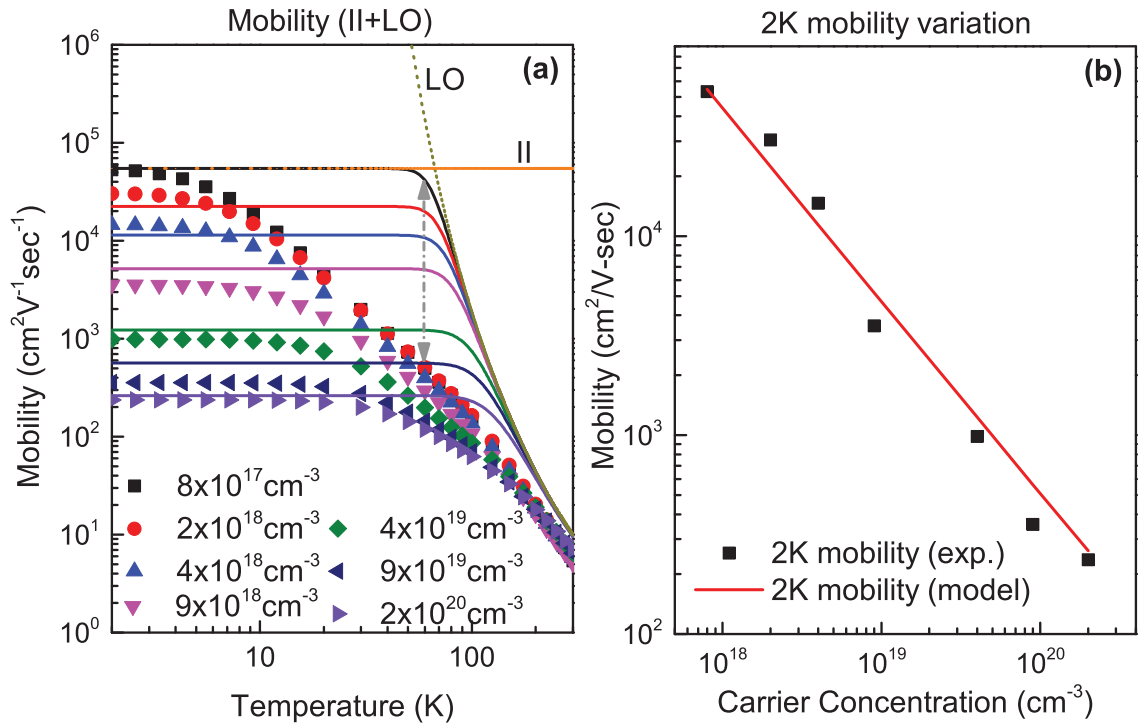

**Fig 1**. a) Temperature dependence of measured (symbols) and modeled (solid lines) electron mobility in SrTiO$_3$ thin films. LO phonon (LO) and ionized impurity (II) scattering is included in the model (shown separately for $n$~$8\times10^{17}$ cm$^{-3}$ sample); the mismatch between theory and experiment for lowest doping sample at ~60K is also shown (grey double-arrow), b) Partial wave expansion model for ionized impurity scattering fit to 2K measured mobility data.

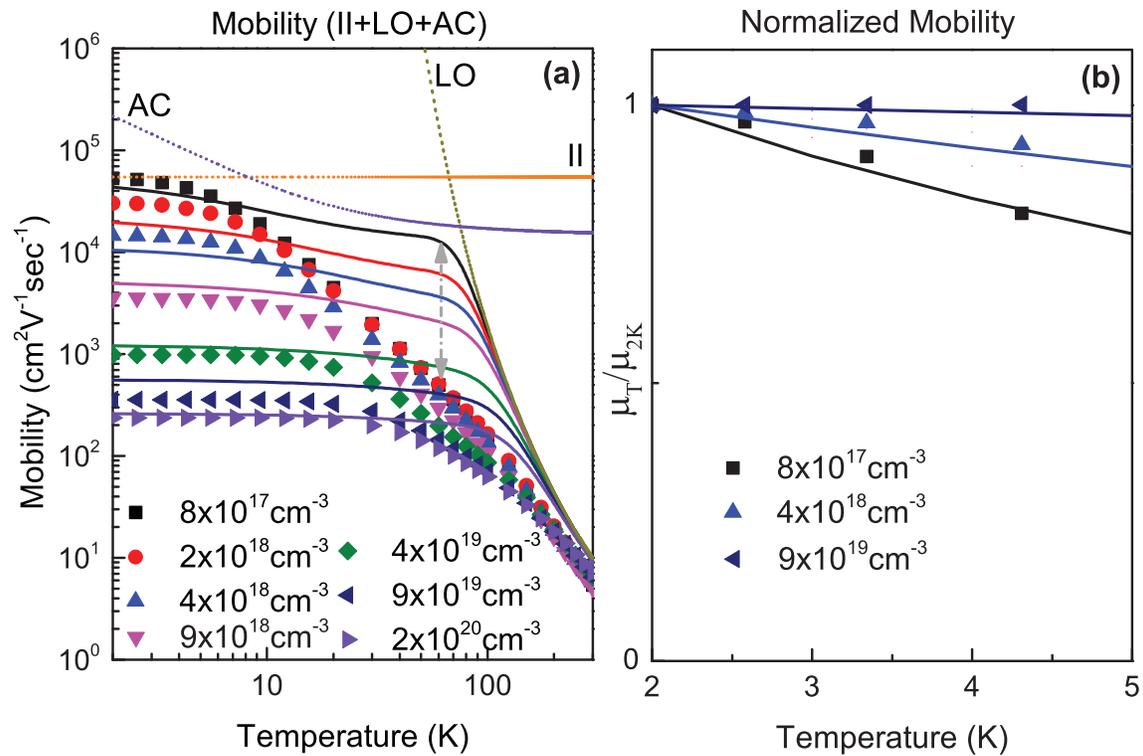

**Fig 2**. a) Temperature dependence of measured (symbols) and modeled (solid lines) electron mobility in SrTiO$_3$ thin films. LO phonon, ionized impurity, and acoustic phonon (AC) scattering is included in the model (shown separately for 8x10$^{17}$ cm$^{-3}$ sample); the mismatch between theory and experiment for lowest doping sample at ~60K is also shown (grey double-arrow) b) Measured and modeled mobility normalized to 2K mobility, showing the agreement in low-temperature slope of model and measurement with inclusion of acoustic phonon scattering.

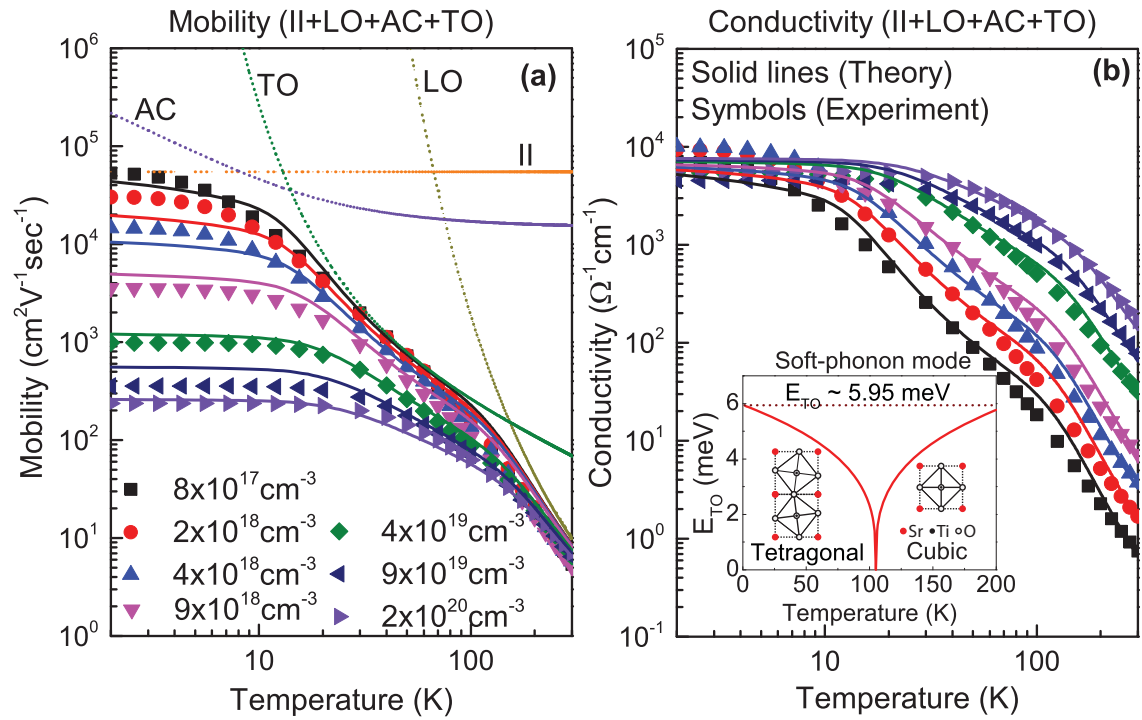

**Fig 3**. (a) Electron mobility, (b) Conductivity of La-doped SrTiO$_3$ thin films. Symbols (measurement), continuous lines (model), dotted lines (individual scattering mechanism for 8x10$^{17}$ cm$^{-3}$ electron conc.). 3b (Inset) 5.95 meV Soft-phonon mode in SrTiO$_3$ responsible for its cubic to tetragonal structural phase transition. Excellent agreement between model and measurement is obtained with inclusion of 5.95 meV TO phonon scattering.